\documentclass[a4paper,10pt]{article}

\usepackage{mathrsfs}

\newcommand{\bfa}[1]{\mbox{\boldmath${#1}$}}

\newcommand{\bea}{\begin{eqnarray}}
\newcommand{\eea}{\end{eqnarray}}
\newcommand{\bnn}{\begin{eqnarray*}}
\newcommand{\enn}{\end{eqnarray*}}
\newcommand{\be}{\begin{equation}}
\newcommand{\ee}{\end{equation}}

\def\PACS{\par\leavevmode\hbox {\it PACS:\ }}%
\def\MSC{\par\leavevmode\hbox {\it MSC:\ }}%
\def\UK{\par\leavevmode\hbox {\it Keywords:\ }}%
\input xy
\xyoption {all}

\begin{document}
\title{Representations of Clifford algebras with hyperbolic numbers}

\author{S. Ulrych\\ Wehrenbachhalde 35, CH-8053 Z\"urich, Switzerland}
\date{August 16, 2007}

\maketitle

\begin{abstract}
The representations of Clifford algebras and their involutions and
anti-involutions are fully investigated since decades. 
However, these representations
do sometimes not comply with usual
conventions within physics. A few simple
examples are presented, which point out that the hyperbolic
numbers can close this gap.
\end{abstract}


{\scriptsize\PACS{03.65.Fd; 02.20.Sv; 02.40.Tt; 02.40.-k; 04.50.+h}
\MSC{81R05; 11E88; 51H30; 37D05; 81V22}
\UK{Clifford algebra; Hyperbolic numbers; Double numbers; Hyperbolic geometry; Complex manifolds}}

\section{Introduction}
The Clifford algebra
approach to physics has become more and more popular over the last decades.
The main idea is to replace the basis vectors
of a real linear space by matrices,
or even to rely purely on the algebraic properties of the 
basis matrices in a form
which is independent of an explicit representation.
The application of Clifford algebras to physics has been strongly promoted 
by Hestenes \cite{Hes66,Hes84,Hes99,Hes03} and has been investigated beside others also
by Gull, Doran, and Lasenby \cite{Gul93,Dor03}, and Rodrigues et al. \cite{Rod05}.

The explicit matrix representations of Clifford 
algebras are well investigated. For a complete summary it is referred 
to the textbooks of Porteous \cite{Por69,Por95}.
Though not mandatory needed, these representations give
additional interesting insights. All representations
of Clifford algebras can be constructed based on the real numbers,
complex numbers, quaternions, and the double fields over these number systems.
Equivalently, and more familiar to physicists, the real numbers, complex numbers,
hyperbolic numbers, and the elements of the Pauli algebra can 
be chosen as basic building blocks (see also Keller \cite{Kel94}).

In the same way as complex numbers can parametrize a circle in two
dimensions, hyperbolic numbers and the double field parametrize a hyperbola
in a plane \cite{Sob95}, where the two number systems refer to the two
possible mathematical representations of a hyperbola. The
hyperbolic numbers are also known as split-complex, paracomplex
or double numbers \cite{Yag79}. 
It is clear that the hyperbolic numbers offer an elegant way
to parametrize the hyperbolic spacetime metric of relativistic
physics. But even more important is the fact that they give the
possibility to consider the general and special linear groups
as unitary groups \cite{Por95,Zho84,Zho85,Zho92}. With this property
the hyperbolic numbers are able to relate concepts from
different areas in physics, namely general relativity and quantum physics.
Furthermore, it has been shown by Hucks \cite{Huc93} that the hyperbolic complex spinor
combines the so-called dotted and undotted spinor representations
of relativistic physics in a single expression, which opens
the way to a new representation of the Dirac theory.

Recently, the hyperbolic 
numbers have been used by da Rocha and Vaz \cite{Roc06}
in order to study chirality in the context of
extended Grassmann and Clifford algebras.
Khrennikov et al. disproved the von Neumann
uniqueness theorem in hyperbolic quantum mechanics \cite{Khr06}.
Boccaletti et al. \cite{Boc07} investigated the twin-paradox
with the help of hyperbolic numbers. 
More references on hyperbolic numbers can be found in \cite{Ulr06Gravi}.

Most applications of Clifford algebras to physics so far
rely on real Clifford algebras. Complex Clifford algebras
have been recently used in the investigation  of the CPT-group
by Varlamov \cite{Var04,Var042}, and in the context of the
Dirac theory by Sabadini et al. \cite{Sab02}, Miralles et al. \cite{Mir01}, 
Marchuk \cite{Mar01}, and
by Avramidi \cite{Avr05}. Polyvector Super-Poincar\'e algebras
have been investigated by Alekseevsky et al. \cite{Ale04}.
In the context of String theory complex Clifford algebras
have been used by Asakawa et al. \cite{Asa02}. They 
complexify with the hyperbolic unit, as well as Moffat in
his Noncommutative Quantum Gravity \cite{Mof00}.

It is one advantage of the explicit representations of Clifford algebras,
based on the building blocks mentioned above, that
the complexification of a real algebra can be performed in
a clear and straightforward way.
One example for such a complexification is based on the
$\bfa{R}_{3,0}$ paravector algebra, which has been considered by
Sobczyk and Baylis \cite{Sob81,Bay89} for the representation of relativistic vectors. 
Baylis has 
shown that the theory of electrodynamics can be fully expressed in terms of
this algebra. In his textbook \cite{Bay99} a wide range of
explicit physical applications of the $\bfa{R}_{3,0}$ algebra can be found.

The algebra $\bfa{R}_{3,0}$ can be represented with the help of the
hyperbolic unit. In this case the algebra will be denoted as hyperbolic algebra, which
has been used recently also by Wei and Xuegang \cite{Wei07}
for the investigation 
of the Lorentz transformation and related special relativistic physics.
The algebra can be complexified further
to provide the complex Clifford algebra $\bar{\bfa{C}}_{3,0}$. It has been 
proposed in \cite{Ulr05pat} to use this algebra to represent physical operators,
like the mass operator, in their most general form. This idea is
reconsidered in this work, which leads to
an investigation of the Clifford algebra $\bfa{R}_{0,5}$
and its complexification.

In order to make this work more self-contained
from a mathematical point of view, a short introduction to
Clifford algebras is presented, which is strongly influenced
by Porteous \cite{Por95}. The notation used in this work 
differs partly from the notation given
in some recent publications \cite{Ulr06Gravi,Ulr05pat}.
The sum convention is not used. The summations
are displayed explicitly. They are split into
two different parts corresponding to the contributions
of a quadratic space with positive and 
negative signature.
All vector coordinates can therefore be written 
with lower indices.

\section{Clifford algebras}
\label{Clialg}
Clifford algebras can be used to represent elements of a quadratic space, i.e.,
a finite-dimensional real linear space with a symmetric
scaler product. The quadratic space with the
scalar product
\be
\label{quad}
(x,y)=-\sum_{1\leq i \leq p}x_i y_i+\sum_{1\leq j \leq q}x_{p+j} y_{p+j}
\ee
will be denoted in the following as $\bfa{R}^{\,p,q}$.
The relationship to Clifford algebras
is given by the quadratic form, which 
corresponds in the context of Clifford algebras to \cite{Por95} 
\be
(x,x)=x\bar{x}\;,
\ee
where the bar symbol indicates the conjugation
anti-involution.

A Clifford algebra can be represented in terms of an orthonomal basis
$\{e_i:1\leq i \leq n\}$, where $n=p+q$. Conjugation reverses the sign of the basis
elements which is leading to (no sum convention)
\be
\label{minus}
e_i\bar{e_i}=-e_i^2\;.
\ee
The basis elements are mutually anticommuting elements of the Clifford algebra
\be
\label{anticom}
e_ie_j+e_je_i=0,\hspace{0.5cm} i\neq j\;,
\ee
and they square to either $0$, $1$, or $-1$. 
The so-called universal Clifford algebra related to the quadratic space $\bfa{R}^{p,q}$ 
will be denoted in the following as $\bfa{R}_{p,q}$ \cite{Por95}. The universal Clifford
algebra is of dimension $2^n$. The real universal Clifford algebras are often
denoted also as $Cl(p,q)$.

\section{Complex and hyperbolic numbers as Clifford algebras}
One can consider the Clifford algebras $\bfa{R}_{\,0,1}$ and $\bfa{R}_{\,1,0}$ 
for the one-dimensional quadratic spaces $\bfa{R}^{\,0,1}$ and $\bfa{R}^{\,1,0}$.
An element of these spaces is expressed in terms of the single basis element in both cases as
$z=xe$.
The quadratic form is then calculated as
\be
z\bar{z} = -x^2 e^2\;.
\ee
If the quadratic space is of signature $(1,0)$ the basis element squares 
according to Eq.~(\ref{quad}) to $e^2=1$.
This basis element will be denoted in the following as the hyperbolic unit $e=j$.
If the quadratic space is of signature $(0,1)$ the basis element squares to $e^2=-1$.
This basis element is denoted as the complex unit $e=i$.
In both cases there are two algebraic distinct elements, the basis element $e$ and the
unity. Both algebras are thus of 
dimension $2^n=2$ and therefore universal Clifford algebras. 
The Clifford algebra $\bfa{R}_{\,0,1}$ is represented by the complex
numbers $\bfa{C}$. A representation of the Clifford algebra $\bfa{R}_{\,1,0}$ is 
given by the hyperbolic numbers over the real number field, denoted here as $\bfa{H}_R$
(Usually $\bfa{H}$ is used to denote the quaternions,
whereas in this text the notation $\bfa{Q}$ is used for the quaternions).
Note, that the
real numbers $\bfa{R}$ are a non-universal Clifford algebra 
of dimension $2^{n-1}=1$ for $\bfa{R}^{\,1,0}$ with
conjugation sending the unity to $-1$.

Beside conjugation, two other involutions play a major role in the
description of Clifford algebras and their structure, graduation and reversion.
Graduation changes the sign of a basis element, whereas reversion reverses the order of
the basis elements in a geometric product.
Involutions that change the order in a geometric product
are denoted as anti-involutions.
In the case of reversion one finds $(ab)^\dagger=b^\dagger a^\dagger$.
Graduation is an involution, which 
does not reverse the order in a product, i.e., $\widehat{ab}=\hat{a}\hat{b}$.
Conjugation, reversion, and graduation are related by $\bar{a}=\hat{a}^\dagger$.
The effect of these involutions on the
hypercomplex units within $\bfa{R}_{\,1,0}$ and $\bfa{R}_{\,0,1}$ is displayed in Table~\ref{invo1dim}.

\begin{table}
\begin{center}
\begin{tabular}{|c|c|c|c|}
\hline
$a$ & $\bar{a}$ & $a^\dagger$ & $\hat{a}$ \\
\hline
$e$ & $-$ & $+$ & $-$\\
\hline
$i$ & $-$ & $+$ & $-$\\
\hline
$j$ & $-$ & $+$ & $-$\\
\hline
\end{tabular}
\end{center}
\caption{Effect of conjugation, reversion, and graduation on the hypercomplex units within $\bfa{R}_{\,1,0}$ and
$\bfa{R}_{\,0,1}$. \label{invo1dim}}
\end{table}
The hyperbolic numbers $\bfa{H}_R$ and the complex
numbers $\bfa{C}$ can be understood also as non-universal Clifford 
algebras of dimension $2^{n-1}=2$ related to the quadratic spaces $\bfa{R}^{\,1,1}$ and $\bfa{R}^{\,0,2}$.
The hyperbolic numbers can be represented in analogy to the complex numbers
\be
\label{realhyp}
z=x+jy\;,\hspace{0.5cm}x,y \in\bfa{R}\;,
\ee
with $j^2=1$.

Instead of the hyperbolic numbers Porteous \cite{Por95} uses the double field,
which he denotes as $^2\bfa{R}$. The double field is the null
basis representation of the hyperbolic number system \cite{Huc93}.
It is obtained if the two basis elements of the algebra are redefined as 
\be
e=\frac{1}{2}(1+j)\;,\hspace{0.5cm}\bar{e}=\frac{1}{2}(1-j)\;.
\ee
One finds $e^2=e$, $\bar{e}^2=\bar{e}$, and $e\bar{e}=0$.
An element of the double field is written as
\be
z=xe+y\bar{e}\;,\hspace{0.5cm}x,y \in\bfa{R}\;,
\ee
which shows also the direct sum structure
of the double field $^2\bfa{R}\cong \bfa{R}\oplus \bfa{R}$.
Alternatively, an element of the double field can be
represented also as a column vector \cite{Por95,Huc93}.
\be
\label{column}
z=(x,y)\;,\hspace{0.5cm}x,y \in\bfa{R}\;.
\ee
Conjugation then corresponds to a swap of the two coordinates. For the
double field the quadratic form is then calculated as
\be
z\bar{z}=xye+yx\bar{e}\equiv (xy,yx)\;,
\ee
whereas for the hyperbolic numbers the quadratic form is given as
\be
z\bar{z}=x^2-y^2\;.
\ee
Points with an equal square form
a hyperbola,
if one understands the hyperbolic numbers and 
the double field as points in the hyperbolic number plane.
The hyperbolic numbers and the double field refer to the
two possible mathematical representations of a hyperbola.
Mathematicians often prefer the representation in terms of the double field,
whereas for physicists the hyperbolic numbers are of special interest because
of their natural relationship to the spacetime metric of relativistic physics.

A concrete application of the hyperbolic numbers to quantum physics
has been given by Khrennikov \cite{Khr03} in order to generalize
the concept of interference of probabilities. The general formula
for the interference of probabilities is given by \cite{Khr03}
\be
\label{proba}
P=P_1+P_2+2\sqrt{P_1P_2}\lambda\;,
\ee
where $P_1$ and $P_2$ are two probabilities and $\lambda\in\bfa{R}$
is a free parameter.
The case $\vert\lambda\vert\leq 1$ is covered by the
parametrization $\lambda = \cos{\theta}$ leading to
complex quantum mechanics. Eq. (\ref{proba}) can
then be linearized in the form
\be
P=\vert \sqrt{P_1} + e^{i\theta}\sqrt{P_2}\vert^2\;,
\ee
which points out the role of the complex numbers for
quantum physics. In general, the case $\vert\lambda\vert\geq 1$ 
needs to be considered for the superposition as well. Here, one
can choose the parametrization $\lambda = \pm\cosh{\theta}$.
With the help of the hyperbolic numbers Eq.~(\ref{proba}) can
be linearized now according to
\be
P=\vert \sqrt{P_1} \pm e^{j\theta}\sqrt{P_2}\vert^2\;,
\ee
where $e^{j\theta}=\cosh{\theta}+j\sinh{\theta}$. This is the 
starting point of hyperbolic quantum mechanics, an approach which has
been investigated in further detail by Khrennikov
\cite{Khr06,Khr03b,Khr04}.

\section{The Clifford algebra $\bfa{R}_{\,3,0}$}
Another example is the universal Clifford algebra $\bfa{R}_{\,3,0}$ for the quadratic
space $\bfa{R}^{\,3,0}$. An element of this algebra can be written as 
\be
z=\sum_{1\leq i \leq 3} x_ie_i\;.
\ee
The quadratic form is then calculated with the help of Eqs.~(\ref{minus}) 
and (\ref{anticom}) as
\be
\label{compsquad}
z\bar{z}=-\sum_{1\leq i \leq 3} x^{2}_ie_i^2\;.
\ee
To obtain the correct signature the square of the basis elements must be equal to  $e_i^2=1$.
Anticommuting elements with this property are provided by the Pauli algebra $\sigma_i$. In the
literature the algebra of complex $2\times 2$ matrices $\bfa{C}(2)$, generated by the
Pauli matrices, is therefore considered as an explicit representation of $\bfa{R}_{\,3,0}$.
However, one has to keep in mind that conjugation has to change the sign of
the basis elements $\bar{e}_i=-e_i$. In order to represent
conjugation as usual with transposition 
and change of sign of the complex and hyperbolic units,
the hyperbolic unit is added to the Pauli matrices. The basis elements of $\bfa{R}_{\,3,0}$
are therefore given as
\be
e_i=j\sigma_i\;.
\ee
Porteous \cite{Por95} does not use the hyperbolic unit to represent $\bfa{R}_{\,3,0}$. 
Instead he defines conjugation of a matrix $a$ in this context as
\be
\label{conpaul}
\bar{a}=\left(\begin{array}{cc}
\;\;\; a_{22}&-a_{12}\\
-a_{21}&\;\;\;a_{11}\\
\end{array}\right)\;,
\ee
which reverses the sign of all Pauli matrices as well.
Though nothing speaks against this representation of the conjugation 
anti-involution from a mathematical point of view,
it is unfamiliar for a physicist.

It should be mentioned that the Pauli algebra multiplied by the hyperbolic unit,
which will be denoted as hyperbolic algebra,
is still isomorphic to $\bfa{C}(2)$.
The only non-trivial expressions that can be generated by multiplication of the basis elements are
$j\sigma_i$, $i\sigma_i$, and $ij$. The algebra is of dimension $2^n=8$ and therefore a universal
Clifford algebra. 
The effect of conjugation, reversion, and graduation on the
hypercomplex units within $\bfa{R}_{\,3,0}$ is displayed in Table \ref{invo}.
Note that this table has been wrongly extended in
recent publications, see e.g. \cite{Ulr06Gravi}, to the
lower dimensional case.

\begin{table}
\begin{center}
\begin{tabular}{|c|c|c|c|}
\hline
$a$ & $\bar{a}$ & $a^\dagger$ & $\hat{a}$ \\
\hline
$e_i$ & $-$ & $+$ & $-$\\
\hline
$\sigma_i$ & $+$ & $+$ & $+$\\
\hline
$i$ & $-$ & $-$ & $+$\\
\hline
$j$ & $-$ & $+$ & $-$\\
\hline
\end{tabular}
\end{center}
\caption{Effect of conjugation, reversion, and graduation on the hypercomplex units within $\bfa{R}_{\,3,0}$
and $\bar{\bfa{C}}_{3,0}$.\label{invo}}
\end{table}
The element with the highest grade within a Clifford algebra is called the pseudoscalar of the algebra.
In this case the element is calculated as
\be
e_1e_2e_3=ij\;.
\ee
The grade corresponds to the number of basis elements that are used to
represent an element of the Clifford algebra. The grade involution distinguishes between
elements of even and odd grade.

Porteous considers the real numbers, complex numbers, quaternions,
and the double fields over these division algebras
as the basic building blocks for the representation of Clifford algebras.
In this work the real numbers, complex numbers
hyperbolic numbers, and the elements of the Pauli algebra are considered
as the basic blocks to obtain a picture that is more familiar to physicists.
The Pauli algebra is related to the quaternions according to 
$q_i=i\sigma_i$ with $q_i\in\bfa{Q}$ denoting the quaternions.
Multiplication of the basis elements of $\bfa{Q}$ provides, beside the unity, only another element within $i\sigma_i$.
Whereas, when starting from $\sigma_i$ one obtains also $i$, and $i\sigma_i$ as additional elements. The quaternions
$\bfa{Q}$, represented within $\bfa{C}(2)$, therefore consist of only one half of the elements 
that can be generated by the
Pauli algebra. 
This relationship is expressed more mathematical by the
fact that the Pauli algebra is the complexification of
the quaternions
\be
\bfa{C}(2)\cong\bfa{Q}\otimes\bfa{C}\;.
\ee
Similar relationships hold also for higher dimensions, e.g.,
between $\bfa{C}(4)$ and $\bfa{Q}(2)$.

\section{The Clifford algebra $\bfa{R}_{\,0,5}$}
\label{fivespace}
Another interesting case is the universal Clifford algebra $\bfa{R}_{\,0,5}$.
This algebra is represented by the $4\times 4$ Pauli matrices
$\bfa{C}(4)$ \cite{Por95}. The quadratic form is given by 
\be
z\bar{z}=-\sum_{1\leq i \leq 5} x^{2}_ie_i^2\;.
\ee
In order to
obtain the correct signature, the square of the basis elements must be equal to $e_i^2=-1$.
The basis elements can be represented as
\be
e_i=i\sigma_{0i}\;,
\ee
where the $4\times 4$ Pauli matrices $\{\sigma_{ab}:0\leq a,b \leq 5\}$ are related
to the generators of $SU(4,\bfa{C})$ by
\be
\label{4gener}
J_{ab}=\frac{\sigma_{ab}}{2}\;.
\ee
The generators satisfy the commutation relations
\be
\label{comrel}
\left[J_{ab},J_{cd}\right]=i(\delta_{ac}J_{bd}-\delta_{ad}J_{bc}-\delta_{bc}J_{ad}
+\delta_{bd}J_{ac})\;.
\ee
The explicit matrix representations of  the $4\times 4$ Pauli matrices are given in Appendix \ref{fivemat}.
The non-trivial elements  that can be obtained by multiplication of the basis elements are
$\sigma_{ab}$, $i\sigma_{ab}$ and $i$. The algebra is thus of dimension $2^n=32$.
The pseudoscalar of the algebra is calculated as
\be
e_1e_2e_3e_4e_5=-i\;.
\ee

Within $\bfa{R}_{\,0,5}$ reversion of the $4\times 4$ matrices is represented
according to Porteous as
\be
a^\dagger=\left(
\begin{array}{cccc}
a_{22}&-a_{12}&a_{42}&-a_{32}\\
-a_{21}&a_{11}&-a_{41}&a_{31}\\
a_{24}&-a_{14}&a_{44}&-a_{34}\\
-a_{23}&a_{13}&-a_{43}&a_{33}\\
\end{array}
\right)\;.
\ee
In order to obtain the relationship $\bar{a}=\hat{a}^\dagger$, where conjugation should
be represented as conjugation and transposition of the matrix,
graduation is introduced as
\be
\hat{a}=\left(
\begin{array}{cccc}
\bar{a}_{22}&-\bar{a}_{21}&\bar{a}_{24}&-\bar{a}_{23}\\
-\bar{a}_{12}&\bar{a}_{11}&-\bar{a}_{14}&\bar{a}_{13}\\
\bar{a}_{42}&-\bar{a}_{41}&\bar{a}_{44}&-\bar{a}_{43}\\
-\bar{a}_{32}&\bar{a}_{31}&-\bar{a}_{34}&\bar{a}_{33}\\
\end{array}
\right)\;.
\ee
Based on these definitions one obtains Table \ref{invo05}, where
$\{e_i,\sigma_{ij}:1\leq i,j \leq 5\}$. 
The elements $i\sigma_{ij}$, which are even under graduation $\hat{a}=a$,
can be used to represent the 
rotations in the quadratic space $\bfa{R}^{\,0,5}$. They 
generate the spin group of $\bfa{R}_{\,0,5}$. The even
subalgebra is isomorphic to $\bfa{R}_{\,0,4}$ and it can be equivalently represented by
$\bfa{Q}(2)$, the quaternionic $2\times 2$ matrices \cite{Por95}.

\begin{table}
\begin{center}
\begin{tabular}{|c|c|c|c|}
\hline
$a$ & $\bar{a}$ & $a^\dagger$ & $\hat{a}$ \\
\hline
$e_i$ & $-$ & $+$ & $-$\\
\hline
$\sigma_{0i}$ & $+$ & $+$ & $+$\\
\hline
$\sigma_{ij}$ & $+$ & $-$ & $-$\\
\hline
$i$ & $-$ & $+$ & $-$\\
\hline
$j$ & $-$ & $+$ & $-$\\
\hline
\end{tabular}
\end{center}
\caption{Effect of conjugation, reversion, and graduation on the used hypercomplex units
within $\bfa{R}_{\,0,5}$ and $\bar{\bfa{H}}_{\!R\,0,5}$. \label{invo05}}
\end{table}

\section{Complex Clifford algebras}
One can distinguish four
different types of complex Clifford algebras \cite{Por95}, i.e., all in all there are five Clifford algebras naturally
associated to an orthogonal space $\bfa{R}^{\,p,q}$. Two complex algebras arise by the
complexification with the complex unit. 
The first of these algebras is related to the standard complex
scalar product
\be
\label{compquad}
(x,y)=\sum_{1\leq i \leq n}x_i y_i\;,\hspace{0.5cm}x,y\in \bfa{C}^n\;,
\ee
with $n=p+q$. This complex Clifford algebra will be denoted as $\bfa{C}_n$.
The other type of complex algebra is related to the Hermitian product
\be
\label{hermite}
(x,y)=-\sum_{1\leq i \leq p}\bar{x}_i y_i+\sum_{1\leq j \leq q}\bar{x}_{p+j} y_{p+j}\;,\hspace{0.5cm}x,y\in \bar{\bfa{C}}^{p,q}\;.
\ee
This algebra will be denoted as $\bar{\bfa{C}}_{p,q}$. 
$\bfa{C}_n$ and $\bar{\bfa{C}}_{p,q}$ are isomorphic,
but with their assigned conjugation, in the first case conjugation 
is the identity, they are not at all the same.
In addition, Porteous considers
complexifications with the double fields over the real and complex numbers,
$^2\bfa{R}^\sigma$ and $^2\bfa{C}^\sigma$, with the swap, indicated by $\sigma$, playing
the role of conjugation as mentioned earlier. Each of the complexified algebras should be
regarded as a superalgebra, that is as a $\bfa{Z}_2$-graded algebra, equipped
with the conjugation anti-involution as an integral part of its structure.

The first example considers the complexification of the algebra $\bfa{R}_{\,1,0}$, which
will be related to the Hermitian product of Eq.~(\ref{hermite}). One finds
\be
\bar{\bfa{C}}_{1,0}\cong\bfa{R}_{\,1,0} \otimes\bar{\bfa{C}}\;.
\ee
The tensor product of algebras refers to a decomposition of an algebra analogous
to the direct sum decomposition of a linear space, but involving
the multiplicative structure rather than the additive structure.
Starting from the hyperbolic number representation given in Eq.~(\ref{realhyp}), the complexification
provides the additional elements $i$ and $ij$. 
The resulting commutative ring will be denoted in this work
as the hyperbolic complex number system
$\bar{\bfa{H}}\equiv \bar{\bfa{H}}_{\bar{C}}$
\be
\label{beg}
z=x+iy+jv+ijw\;,\hspace{0.5cm}x,y,v,w \in\bfa{R}\;,
\ee
which is isomorphic to $^2\bar{\bfa{C}}^\sigma$ in \cite{Por95}. 
The same number system can be obtained 
by complexification of the algebra $\bfa{R}_{\,0,1}$ with the hyperbolic unit
\be
\bar{\bfa{H}}_{\!R\,0,1}\cong\bfa{R}_{\,0,1}\otimes\bar{\bfa{H}}_R\;.
\ee
This algebra corresponds to $^2\bfa{R}^\sigma_{\,0,1}$ in the notation of Porteous.

The universal Clifford algebra $\bfa{R}_{\,3,0}$ can be complexified with either the complex 
or the hyperbolic unit
\be
\bar{\bfa{C}}_{\,3,0}\cong\bfa{R}_{\,3,0} \otimes\bar{\bfa{C}}\cong\bfa{R}_{\,3,0}\otimes\bar{\bfa{H}}_R\cong\bar{\bfa{H}}_{\!R\,3,0}\;.
\ee
Based on the representation of the
algebra given in the last section, the complexification 
provides the additional elements $i$, $j$, $\sigma_i$, and $ij\sigma_i$.
The matrix representation of the algebra is thus
given by $\bfa{H}(2)$, the hyperbolic complex
$2\times 2$ matrices (see also Corollary 15.30 in \cite{Por95}).

The complexification of the algebra $\bfa{R}_{\,0,5}$ with the hyperbolic unit
is denoted as
\be
\bar{\bfa{H}}_{\!R\,0,5}\cong\bfa{R}_{\,0,5} \otimes\bar{\bfa{H}}_R\;.
\ee
This algebra corresponds to $^2\bfa{R}^\sigma_{\,0,5}$ in the notation of Porteous.
The complexification provides the additional elements $j$, $ij$, $j\sigma_{ab}$, and $ij\sigma_{ab}$,
i.e., the algebra has all in all sixty-four elements.
The algebra is represented in terms of the hyperbolic complex $4\times 4$ matrices
$\bfa{H}(4)$.

\section{Rotations}
\label{rot}
For an element of a universal Clifford algebra
orthogonal transformations, namely rotations and anti-rotations, are
represented according to Porteous \cite{Por95} as
\be
z\mapsto gz\hat{g}^{-1}\;,
\ee
where $g$ is an element of the Clifford group, whose elements
can be represented as 
the product of a finite number of elements of
the Clifford algebra. For an even number of elements
the grade involution results in
$\hat{g}=g$, i.e.
\be
z\mapsto gzg^{-1}\;,
\ee
which corresponds to a rotation. This reflects the fact that rotations can be represented as
an even number of hyperplane reflections. The odd $\hat{g}=-g$ elements lead to
\be
z\mapsto -gzg^{-1}\;,
\ee
which corresponds to an anti-rotation. However, these definitions 
are only valid for
universal Clifford algebras. In this work essentially paravector
algebras are considered, which leads to changes compared
to the standard picture.

The space of paravectors  is given as $Y=\bfa{R}\oplus  X$, where $X$ refers to a linear space,
which is represented by the Clifford algebra $A$.
The orthogonal transformations that leave the 
quadratic form of a paravector invariant are 
given according to Corollary 16.10 in \cite{Por95} as
the set $\Omega$, which is defined as 
\be
\label{rotadef}
\Omega=\{g\in A:z\in Y\Rightarrow gz\hat{g}^{-1}\in Y\}\;.
\ee

\section{The quasi-sphere in the one-dimensional hyperbolic complex space}
To give a more detailed view
on the rotations defined in the last section
the real linear space $\bfa{R}\oplus\bfa{R}^{\,2,1}\cong \bfa{R}^{\,2,2}$ is considered.
The hyperbolic complex numbers $\bar{\bfa{H}}\cong \bar{\bfa{C}}_{\,1,0}$ equipped with conjugation 
as involution can be used as a non-universal 
paravector algebra for this quadratic space.
The algebra $A$ and the space of paravectors
$Y$ can be represented by $\bar{\bfa{C}}_{\,1,0}$. 
The set $\Omega$, which is used to represent rotations and anti-rotations,
then corresponds to the invertible elements of $\bar{\bfa{C}}_{\,1,0}$.

Conjugation of the complex and the hyperbolic unit is defined as in Table \ref{invo1dim}.
The square of the hyperbolic number is calculated as
\be
\label{square}
z\bar{z}=x^2+y^2-v^2-w^2+2ij(xw-yv)\;.
\ee
In the following, only a reduced subset of $\Omega$ is considered
in order to get some additional insights into the structure of Eq.~(\ref{square}).
Therefore, only rotations
obeying the condition $g\bar{g}=1$, which are identified with the
elements of the spin group, are used in the following investigation.
They can be written in the form
\be
\label{rotgroup}
g=\exp{(- i\phi /2 + j\xi /2)}\;.
\ee
Based on these conditions one finds
\be
z\mapsto gz\hat{g}^{-1}=gzg=zg^2\;.
\ee

The elements $g$ given in Eq.~(\ref{rotgroup}) represent the
one-dimensional hyperbolic complex quasi-sphere
\be
\label{onedimh}
\mathscr{S}(\bar{\bfa{H}})=\{z\in \bar{\bfa{H}}:z\bar{z}=1\}\;.
\ee
The quasi-sphere is isomorphic to
$\mathscr{S}(\bar{\bfa{H}})\cong  \bfa{R}\times S^1$,
which is in contrast to the quasi-spheres 
$\mathscr{S}(\bfa{R}^{\,2,2})\cong \mathscr{S}(\bar{\bfa{C}}^{\,1,1})\cong \bfa{R}^{\,2}\times S^1$. 
The quasi-sphere $\mathscr{S}(\bar{\bfa{H}})$ 
is the group manifold of the hyperbolic unitary group $U(1,\bfa{H})$, which
is locally isomorphic to $U(1,\bfa{C})\times U(1,\bfa{C})$.
The generators of the hyperbolic unitary group
can be identified as $J=1/2$ and $K=ij/2$, and thus
the generators of $U(1,\bfa{C})\times U(1,\bfa{C})$
may be defined as
\be
\label{isomorph}
A=\frac{1}{2}(J+ijK)\;,\hspace{0.5cm}B=\frac{1}{2}(J-ijK)\;.
\ee

The rotation given in Eq.~(\ref{rotgroup}) is an element of the spin group.
As mentioned by Hucks \cite{Huc93} the hyperbolic
representation combines the transformations acting on the 
so-called dotted and undotted spinors.
In the null basis representation the transformation is given as
\be
\label{nullrot}
g=\left(\,\exp{(-i\phi/2)}\exp{(\xi/2)}\,,\,\exp{(-i\phi/2)}\exp{(-\xi/2)}\,\right),
\ee
where the notation of Eq.~(\ref{column}) has been used.
The corresponding quasi-sphere is denoted as $\mathscr{S}(^2\bar{\bfa{C}}^\sigma)\cong\mathscr{S}(\bar{\bfa{H}})$.

An application in quantum physics is obtained
if $U(1,\bfa{H})$, with $\mathscr{S}(\bar{\bfa{H}})$
as the corresponding group manifold, is interpreted 
as a charge symmetry, namely as a generalization of
the electromagnetic charge symmetry. 
It has been originally proposed by
Moffat \cite{Mof82} to understand the hyperbolic unit as the fermion charge in analogy
to the electric charge represented by the complex unit. 
In \cite{Ulr06Gravi} this fermion charge has
been used to introduce a Maxwell theory of gravitation.
Consider in this context
also the recent work of Notte-Cuello and Rodrigues \cite{Not06},
who represent gravitational theory in a Maxwell-like form.

\section{Minkowski space}
The Dirac algebra
is normally considered as the underlying 
Clifford algebra for the 
Minkowksi space $\bfa{R}\oplus\bfa{R}^{\,3,0}\cong \bfa{R}^{\,3,1}$.
However, it is also possible to represent the Minkowski space
in terms of the algebra $\bfa{R}_{\,3,0}$.
This approach has been successfully applied to electrodynamics
by Baylis \cite{Bay99}.

The corresponding paravector algebra of 
$\bfa{R}_{\,3,0}$ can be represented also
in terms of the hyperbolic algebra
\be
\label{base}
 e_a=(1,j\sigma_i)\;,
\ee
and the Minkowski vector $x_a=(x_0,x_i)\in\bfa{R}^{\,3,1}$ is given in this basis as
the paravector
\be
\label{veco}
x=\sum_{0\leq a \leq 3} x_a e_a\;.
\ee
The pseudoscalar of the
hyperbolic algebra can be defined in analogy to Baylis \cite{Bay99} as 
\be
e_0\bar{e}_1e_2\bar{e}_3=ij\;.
\ee
In this context it is useful to introduce the dot and the wedge products. The dot product
corresponds to
\be
\label{scalar}
x\cdot y
=\frac{1}{2}(x\bar{y}+y\bar{x})\;.
\ee 
The wedge product is given as
\be
x\wedge y
=\frac{1}{2}(x\bar{y}-y\bar{x})\;.
\ee
More details about the wedge product can be found in Appendix \ref{wedge}.
The basis elements form a non-cartesian orthogonal basis with respect to the
scalar product
\be
 e_{a}\cdot e_{b}=g_{ab}\;,
\ee
where $g_{ab}$ is the metric tensor of the Minkowski space.
The rotations are given again as
\be
x\mapsto gx\hat{g}^{-1}\;,
\ee
where $g$ is an element of $\Omega$ with the additional
restrictions introduced in Section \ref{rot}. It can be represented as
\be
\label{3dim}
g=\exp{(-i\bfa{\phi} /2 + j\bfa{\xi} /2)}\;,
\ee
where
\be
\label{rotrep}
\bfa{\phi}=\sum_{1\leq i\leq 3}\phi_i \sigma_i\;,\hspace{0.5cm}\bfa{\xi}=\sum_{1\leq i\leq 3}\xi_i \sigma_i\;.
\ee
It is important to note that due to Table \ref{invo} the effect of $\hat{g}^{-1}$ is different for
the complex and the hyperbolic part of the transformation. For a pure complex 
transformation one finds $\hat{g}=g$ and for a pure hyperbolic transformation, 
which represents a boost, one obtains $\hat{g}=g^{-1}$.
In both cases there is the relationship $\hat{g}^{-1}=g^\dagger$.

The generators of the transformation
can be identified as $J_i=\sigma_i/2$ and $K_i=ij\sigma_i/2$.
The commutation relations correspond to the commutation relations of
the Lorentz group. For rotations one finds
\be
[J_i,J_j]=i\epsilon_{ijk}J_k\;,
\ee
for boosts
\be
[K_i,K_j]=-i\epsilon_{ijk}K_k\;,
\ee
and for the mixing the commutation relations are
\be
[J_i,K_j]=i\epsilon_{ijk}K_k\;.
\ee

The operators generate the group $SU(2,\bfa{H})$, 
which can be identified with the spin group of $SO(3,1,\bfa{R})$.
An isomorphism to the group $SU(2,\bfa{C})\times SU(2,\bfa{C})$ can be constructed
in analogy to Eq.~(\ref{isomorph}). 
\be
A_i=\frac{1}{2}(J_i+ijK_i)\;,\hspace{0.5cm}B_i=\frac{1}{2}(J_i-ijK_i)\;.
\ee
Because of the known isomorphism to the special linear group,
it follows also that $SU(2,\bfa{H})\cong SL(2,\bfa{C})$.
This isomorphism can be extended to the groups $U(2,\bfa{H})\cong GL(2,\bfa{C})$ (compare with Zhong \cite{Zho84,Zho85,Zho92}).

\section{The six-dimensional Euclidean space}
In this section the paravector algebra for the quadratic space
$\bfa{R}\oplus\bfa{R}^{\,0,5}\cong \bfa{R}^{\,6}$ is investigated.
One can work here with the algebra $\bfa{R}_{\,0,5}$ introduced in Section \ref{fivespace} \cite{Por95}.
The basis elements of the quadratic space are given as
\be
e_a=(1,i\sigma_{0i})\;,
\ee
where $\{e_a:0\leq a \leq 5\}$ and $\{\sigma_{0i}:1\leq i \leq 5\}$. 
The rotations can be represented with the generators of the group $SU(4,\bfa{C})$
as 
\be
x\mapsto gx\hat{g}^{-1}\;,
\ee
where
\be
x=\sum_{0\leq a \leq 5} x_a e_a\;,
\ee
and $g$ is again an element of $\Omega$ with the additional
restrictions introduced in Section \ref{rot}. The rotation can be represented as
\be
\label{5dim}
g=\exp{(-i\bfa{\phi} /2})\;,
\ee
where
\be
\bfa{\phi}=\frac{1}{2}\sum_{0\leq a,b\leq 5}\phi_{ab} \sigma_{ab}\;.
\ee
It is interesting to derive a parameter representation of the sphere
$S^5$ with radius $r$ starting from a basis vector of the form 
$x_a=(0,0,0,0,0,r)$.
The rotation matrix $g$ is given by the five parameter rotation
\be
\label{rota}
g=g_5g_4g_3g_2g_1\;,
\ee
where the first rotation corresponds to a rotation in the plane spanned
by the dimensions $2$ and $5$. The second rotation
is done in the plane of dimension $0$ and $2$, and the following
rotations are performed in the planes $0-1$, $3-5$, and $3-4$.
The explicit form of the rotations is given as
\bea
\label{exp06rot}
g_1&=&\exp{(i\phi_{25} \sigma_{25}/2)}\; ,\nonumber\\
g_2&=&\exp{(-i\phi_{02}\sigma_{02}/2)}\;,\nonumber\\
g_3&=&\exp{(i\phi_{01}\sigma_{01}/2 )}\;,\nonumber\\
g_4&=&\exp{(i\phi_{35}\sigma_{35}/2)}\;,\nonumber\\
g_5&=&\exp{(-i\phi_{34}\sigma_{34}/2)}\;.
\eea

During the calculation one should keep in mind that the 
group elements $\hat{g}^{-1}_2=g_2$
and $\hat{g}^{-1}_3=g_3$ are invariant under the combined action of graduation and inversion
according to Table \ref{invo05}.
Again there is the relationship $\hat{g}^{-1}=g^\dagger$
for all transformations.
One then obtains the parametrization of the sphere with radius $r$ 
\be
\label{sphrep}
x=\left(\begin{array}{c}
r\sin{\phi_{25}}\sin{\phi_{02}}\cos{\phi_{01}}\\
r\sin{\phi_{25}}\sin{\phi_{02}}\sin{\phi_{01}}\\
r\sin{\phi_{25}}\cos{\phi_{02}}\\
r\cos{\phi_{25}}\sin{\phi_{35} }\cos{\phi_{34}}\\
r\cos{\phi_{25}}\sin{\phi_{35} }\sin{\phi_{34}}\\
r\cos{\phi_{25}}\cos{\phi_{35}}\\
\end{array}\right)\;.
\ee
 
\section{Hyperbolic complex extension to the linear space $\bfa{R}^{\,6,6}$}
The hyperbolic unitary group $SU(4,\bfa{H})$ is obtained as a generalization of 
the complex group $SU(4,\bfa{C})$ if the generators given in Eq.~(\ref{4gener}) are multipled
by the hyperbolic complex pseudoscalar
\be
K_{ab}=ijJ_{ab}\;.
\ee
The commutation relations for the thirty parameter group
are then extended by
\be
\left[J_{ab},K_{cd}\right]=i(\delta_{ac}K_{bd}-\delta_{ad}K_{bc}-\delta_{bc}K_{ad}
+\delta_{bd}K_{ac})\;
\ee
and
\be
\left[K_{ab},K_{cd}\right]=-i(\delta_{ac}K_{bd}-\delta_{ad}K_{bc}-\delta_{bc}K_{ad}
+\delta_{bd}K_{ac})\;.
\ee
Note, that this group is locally isomorphic to the special linear group $SL(4,\bfa{C})$ in analogy
to the isomorphism between the relativistic spin groups
$SL(2,\bfa{C})$ and $SU(2,\bfa{H})$. This isomorphism is again obtained
via the generators of the group $SU(4,\bfa{C})\times SU(4,\bfa{C})$
\be
\label{10gen}
A_{ab}=\frac{1}{2}(J_{ab}+ijK_{ab})\;,\hspace{0.5cm}B_{ab}=\frac{1}{2}(J_{ab}-ijK_{ab})\;.
\ee
One can construct now a non-universal paravector algebra for the
quadratic space $\bfa{R}^{\,6,6}$ with the basis vector
\be
\label{twelve}
e_a=(1,i\sigma_{0i},ij,-j\sigma_{0i})\;,
\ee
where $\{e_a:0\leq a \leq 11\}$.
The rotations are extended to
\be
\label{10dim}
g=\exp{(-i\bfa{\phi} /2}+ j\bfa{\xi}/2)\;,
\ee
with the additional transformations
\be
\bfa{\xi}=\frac{1}{2}\sum_{0\leq a,b\leq 5}\xi_{ab} \sigma_{ab}\;.
\ee

The extended set of transformations forms the
quasi-sphere $\mathscr{S}(\bar{\bfa{H}}^{\,3})$, which can be represented
in terms of rotations in five $\bfa{R}^{\,2,2}$ subspaces.
The transformations are represented as generalizations of the rotations
in Eq.~(\ref{exp06rot})
\be
\phi_{ab}\rightarrow\tilde{\phi}_{ab}=\phi_{ab}+ij\xi_{ab}\;.
\ee
The subsphere related to $g_1$ is defined in the $\bfa{R}^{\,2,2}$ subspace formed
by the dimensions with number $2$, $5$, $8$, and $11$. Similar generalizations
hold for the other rotations.
The parametrization of the sphere in Eq.~(\ref{sphrep}) generalizes to the angles
$\tilde{\phi}_{ab}$.
The explicit form of the quasi-sphere 
$\mathscr{S}(\bar{\bfa{H}}^{\,3})\cong \bfa{R}^{\,5}\times S^5$ can be obtained by the relations
\be
\cos{\tilde{\phi}_{ab}}=\cos{\phi_{ab}}\cosh{\xi_{ab}}-ij\sin{\phi_{ab}}\sinh{\xi_{ab}}\;,
\ee
and
\be
\sin{\tilde{\phi}_{ab}}=\sin{\phi_{ab}}\cosh{\xi_{ab}}+ij\cos{\phi_{ab}}\sinh{\xi_{ab}}\;.
\ee
The contributions proportional to $ij$ in each subspace 
$\{e_a:0\leq a \leq 5\}$ are assigned to 
the dimension $e_{a+6}$ in the full twelve-dimensional space. 
The explicit result of this
calculation does not provide any further information.

For further investigations in the context of
quantum physics one can consider now the 
spin representation of $SU(4,\bfa{H})$ related to 
the transformations in Eq.~(\ref{10dim}), the 
$SU(4,\bfa{C})\times SU(4,\bfa{C})$ representation given by the
generators of Eq.~(\ref{10gen}), or the null basis
representation
\be
\label{10nullrot}
g=\left(\,\exp{(-i\bfa{\phi}/2)}\exp{(\bfa{\xi}/2)}\,,\,\exp{(-i\bfa{\phi}/2)}\exp{(-\bfa{\xi}/2)}\,\right),
\ee
according to Eq.~(\ref{nullrot}), which reveals the
dotted and undotted spinor structure.

\section{Application to quantum physics}
\label{pati}
The concepts introduced in the last sections can be 
applied to quantum physics. 
In a series of papers \cite{Ulr06Gravi,Ulr05pat,Ulr06Spin}
it has been proposed to use the so-called mass operator equation as the
fundamental fermion equation of relativistic quantum physics as an alternative to
the Dirac equation. The mass operator takes the form
\be
M^2=p\bar{p}\;,
\ee
where $p\in\bfa{R}^{\,3,1}$ is an element of the Minkowski momentum space.

It has been suggested in \cite{Ulr05pat} to generalize the momentum
operator to the complex Clifford algebra $\bar{\bfa{C}}_{\,3,0}$,
which has the consequence that the momentum operator
is extended to the hyperbolic complex 
paravector space $p\in\bar{\bfa{H}}^{\,3,1}$
\be
\label{begmom}
p=q+io+js+iju\;,\hspace{0.5cm}q,o,s,u \in\bfa{R}^{\,3,1}\;.
\ee
The momentum vector can thus be considered as an
element of a sixteen-dimensional real vector space.
The model proposed in \cite{Ulr05pat} follows 
a top-down approach. The
total space is given as the starting point, with the goal to derive the internal
particle symmetries from this space. 
(From a philosophical point of view 
such an approach has been suggested
also by Auyang \cite{Auy95}).

The generalized mass operator is expected to be Hermitian.
Hermitecity of an operator $A$ in the presence of the hyperbolic
unit is defined by the condition $\bar{A}=A$ and further with the
stronger requirement that all eigenvalues must be real numbers.
In the language of this work this means that the mass operator
must describe a quasi-sphere
$\mathscr{S}(\bar{\bfa{H}}^{\,3,1})$ with radius $m$ in momentum space,
where $m$ denotes the particle mass.

As a standard momentum vector, which is part of this
quasi-sphere, one may choose the
real vector $p=(m,0)$. The stabilizer of this vector is
acting on the coordinates $q_i$, $o_i$, $s_i$, and $u_i$.
They can be organized into a twelve-dimensional vector
\be
p^f_a=(q_i, o_i, s_i, u_i)\;,
\ee
where $\{p^f_a:0\leq a \leq 11\}$ and $\{q_i, o_i, s_i, u_i:1\leq i \leq 3\}$.
In Eq.~(\ref{begmom}) the coordinates $s_i$ and $u_i$ are multiplied  
by the hyperbolic unit and therefore the vector is identified
as an element of the space $\bfa{R}^{\,6,6}$.

The algebra $\bar{\bfa{C}}_{\,3,0}$ is considered as the base algebra, which
defines also the quasi-sphere structure of the generalized momentum space. However,
if the momentum is an element of a
sixteen-dimensional real vector space this has the consequence that
the possible transformations acting on the momentum
vector go beyond the algebra $\bar{\bfa{C}}_{\,3,0}$. Implicitly, the
algebraic structure is extended.
In this sense, the elements $e_a$ of Eq.~(\ref{twelve})
may be chosen as the basis for the vector $p^f$.
One can consider now the group 
$SO(6,\bfa{R})\times SO(6,\bfa{R})$ as the stabilizer
acting on the vector coordinates, or the spin group
$SU(4,\bfa{H})$ acting on the basis matrices of the 
Clifford algebra.

A principle bundle can be introduced
with the quasi-sphere $\mathscr{S}(\bfa{R}^{\,3,1})$ of radius
$m$ as the base space, which is generated by the Lorentz transformations
acting on the standard vector, and the stabilizer group as its fibre.
This principal bundle satisfies the restriction given
by the hermitecity of the mass operator.
As shown in the last section, the stabilizer is isomorphic
to the group $SU(4,\bfa{C})\times SU(4,\bfa{C})$, which corresponds to
the gauge group of
the Pati-Salam model \cite{Pat73}. 
In addition, Casalbuoni et al. \cite{Cas81} proposed a subconstituent model, in which the group
$U(4,\bfa{C})\times U(4,\bfa{C})$ is used 
as the classification group for subconstituent 
particles and anti-particles in order to enhance the Harari-Shupe 
model \cite{Har79,Shu79}.

The main idea behind the model 
proposed in \cite{Ulr05pat} can thus be summarized as follows:
in the same way
as the underlying Clifford algebra of Minkowski spacetime with the
spin group $SU(2,\bfa{H})$
shows up in the particle spin, the underlying complex Clifford algebra
of the fibre space $\bfa{R}^{\,6,6}$ with the spin group $SU(4,\bfa{H})$ appears to be 
visible in the isospin structure of the subconstituent particle.


\section{Summary}
Mathematicians prefer to represent Clifford algebras
based on the fundamental division algebras of real numbers,
complex numbers, quaternions, and the double fields
over the corresponding division alegebras
$^2\bfa{R}\cong \bfa{R}\oplus \bfa{R}$, $^2\bfa{C}\cong \bfa{C}\oplus \bfa{C}$,
and $^2\bfa{Q}\cong \bfa{Q}\oplus \bfa{Q}$. This provides a
consistent and well founded framework, which can cover all mathematical
structures in this context.

Physicists have to relate mathematics to structures
suitable to represent relativistic spacetime and
quantum physics. For them it might be more convenient
to repesent real and complex Clifford algebras
in terms of the hyperbolic
numbers, the real numbers, the complex numbers and
the Pauli matrices. It has been shown in some concrete
examples how these representations can be introduced.

\appendix
\section{The $4\times 4$ Pauli matrices}
\label{fivemat}
In order to understand the relationship between the $4\times 4$ Pauli matrices,
given as a tensor product of the $2\times 2$ Pauli matrices, and their role
as generators of the group $SU(4,\bfa{C})$, the matrices are labelled in two different
ways. In relationship with the tensor product
\be
A\otimes B=
\left(\begin{array}{cccc}
a_{11}b_{11}&a_{11}b_{12}&a_{12}b_{11}&a_{12}b_{12}\\
a_{11}b_{21}&a_{11}b_{22}&a_{12}b_{21}&a_{12}b_{22}\\
a_{21}b_{11}&a_{21}b_{12}&a_{22}b_{11}&a_{22}b_{12}\\
a_{21}b_{21}&a_{21}b_{22}&a_{22}b_{21}&a_{22}b_{22}\\
\end{array}\right)\;
\ee
the labelling of the $4\times 4$ Pauli matrices is given as
$\{\sigma_{a}:1\leq a \leq 15\}$. 
The first group of matrices is calculated as
\be
\sigma_i\otimes 1\;,
\ee
where in the tensor product the notation
$\{\sigma_{i}:1\leq i \leq 3\}$ refers 
to the $2\times 2$ Pauli matrices.
With the tensor product one can calculate the
$4\times 4$ matrix representations
\be
\sigma_{1}=
\left(\begin{array}{cccc}
0&0&1&0\\
0&0&0&1\\
1&0&0&0\\
0&1&0&0\\
\end{array}\right)\;,
\ee
\be
\sigma_{2}=
\left(\begin{array}{cccc}
0&0&-i&0\\
0&0&0&-i\\
i&0&0&0\\
0&i&0&0\\
\end{array}\right)\;,
\ee
\be
\sigma_{3}=
\left(\begin{array}{cccc}
1&0&0&0\\
0&1&0&0\\
0&0&-1&0\\
0&0&0&-1\\
\end{array}\right)\;.
\ee
The matrices $1$ and $3$ are of special interest, because these matrices,
multiplied by the complex unit, appear as basis elements of the
$\bfa{R}_{\;0,5}$ Clifford algebra. The matrices $4$, $5$, and $6$ are calculated as
\be
1\otimes \sigma_i\;.
\ee
\be
\sigma_{4}=
\left(\begin{array}{cccc}
0&1&0&0\\
1&0&0&0\\
0&0&0&1\\
0&0&1&0\\
\end{array}\right)\;,
\ee
\be
\sigma_{5}=
\left(\begin{array}{cccc}
0&-i&0&0\\
i&0&0&0\\
0&0&0&-i\\
0&0&i&0\\
\end{array}\right)\;,
\ee
\be
\sigma_{6}=
\left(\begin{array}{cccc}
1&0&0&0\\
0&-1&0&0\\
0&0&1&0\\
0&0&0&-1\\
\end{array}\right)\;.
\ee
The third group is related to
\be
\sigma_1\otimes \sigma_i\;.
\ee
Again it should be mentioned that the $2\times 2$ Pauli matrices appear in the
tensor product,
whereas the result is a $4\times 4$ matrix
\be
\sigma_{7}=
\left(\begin{array}{cccc}
0&0&0&1\\
0&0&1&0\\
0&1&0&0\\
1&0&0&0\\
\end{array}\right)\;,
\ee
\be
\sigma_{8}=
\left(\begin{array}{cccc}
0&0&0&-i\\
0&0&i&0\\
0&-i&0&0\\
i&0&0&0\\
\end{array}\right)\;,
\ee
\be
\sigma_{9}=
\left(\begin{array}{cccc}
0&0&1&0\\
0&0&0&-1\\
1&0&0&0\\
0&-1&0&0\\
\end{array}\right)\;.
\ee
The fourth group is related to
\be
\sigma_2\otimes \sigma_i\;.
\ee
This group is again of special interest, because these matrices,
multiplied by the complex unit, correspond to the
remaining three basis elements of $\bfa{R}_{\;0,5}$ 
\be
\sigma_{10}=
\left(\begin{array}{cccc}
0&0&0&-i\\
0&0&-i&0\\
0&i&0&0\\
i&0&0&0\\
\end{array}\right)\;,
\ee
\be
\sigma_{11}=
\left(\begin{array}{cccc}
0&0&0&-1\\
0&0&1&0\\
0&1&0&0\\
-1&0&0&0\\
\end{array}\right)\;,
\ee
\be
\sigma_{12}=
\left(\begin{array}{cccc}
0&0&-i&0\\
0&0&0&i\\
i&0&0&0\\
0&-i&0&0\\
\end{array}\right)\;.
\ee
Finally the fifth group is related to
\be
\sigma_3\otimes \sigma_i\;,
\ee
and the remaining matrices are calculated as
\be
\sigma_{13}=
\left(\begin{array}{cccc}
0&1&0&0\\
1&0&0&0\\
0&0&0&-1\\
0&0&-1&0\\
\end{array}\right)\;,
\ee
\be
\sigma_{14}=
\left(\begin{array}{cccc}
0&-i&0&0\\
i&0&0&0\\
0&0&0&i\\
0&0&-i&0\\
\end{array}\right)\;,
\ee
\be
\sigma_{15}=
\left(\begin{array}{cccc}
1&0&0&0\\
0&-1&0&0\\
0&0&-1&0\\
0&0&0&1\\
\end{array}\right)\;.
\ee

The above fifteen matrices will be set in relationship to
the group generators $\{\sigma_{ab}:0\leq a,b \leq 5\}$ of $SU(4,\bfa{C})$
with the following matrix
\be
\sigma_{ab}=\left(
\begin{array}{cccccc}
0&\sigma_{1}&-\sigma_{3}&\sigma_{10}&\sigma_{11}&\sigma_{12}\\
-\sigma_{1}&0&\sigma_{2}&\sigma_{13}&\sigma_{14}&\sigma_{15}\\
\sigma_{3}&-\sigma_{2}&0&\sigma_{7}&\sigma_{8}&\sigma_{9}\\
-\sigma_{10}&-\sigma_{13}&-\sigma_{7}&0&\sigma_{6}&-\sigma_{5}\\
-\sigma_{11}&-\sigma_{14}&-\sigma_{8}&-\sigma_{6}&0&\sigma_{4}\\
-\sigma_{12}&-\sigma_{15}&-\sigma_{9}&\sigma_{5}&-\sigma_{4}&0\\
\end{array}
\right)\;.
\ee
The matrices in the first row, multiplied by the complex unit, form the
basis elements of the Clifford algebra $\bfa{R}_{\,0,5}$.
The four matrices in the second row on the right 
side of the zero, multiplied by the complex unit, form the
basis elements of the Clifford algebra $\bfa{R}_{\,0,4}$.

\section{Wedge products}
\label{wedge}
The wedge product within the hyperbolic algebra $\bfa{R}_{\,3,0}$ is defined
for a product of two paravectors, which are elements of
the Minkowski space $x,y\in \bfa{R}^{\,3,1}$, as
\be
x\wedge y
=\frac{1}{2}(x\bar{y}-y\bar{x})\;.
\ee
The wedge
product tranforms the paravectors $x$ and $y$ into a
biparavector \cite{Bay99}.
The wedge product can be extended also
to three paravectors
\bea
x\wedge y\wedge v&=&\frac{1}{3!}(x\bar{y} v+y \bar{v} x
+v \bar{x} y \\
&&-y \bar{x} v-x \bar{v} y -v\bar{y} x )\nonumber\;,
\eea
which corresponds to a triparavector.
Finally, a product of four paravectors results in a pseudoscalar
\be
x\wedge y\wedge v\wedge w = \frac{1}{4!}(x\bar{y} v\bar{w}+ 23\mathrm{\;permutations})\;.
\ee
These wedge products have been denoted in \cite{Ulr06Spin} by the bracket
notation $\langle x\bar{y}\rangle_-$ to indicate the anti-symmetry. It is recommended
to use the notation given here, because it corresponds to the common standard.

\end{document}